# Revisiting the Gaia Hypothesis: Maximum Entropy, Kauffman's 'Fourth Law' and Physiosemeiosis

by

Carsten Herrmann-Pillath

East-West Centre for Business Studies and Cultural Science
Frankfurt School of Finance and Management
Sonnemannstraße 9-11
60314 Frankfurt am Main, Germany
Email c.herrmann-pillath@fs.de

**Abstract**

Recently, Kleidon suggested a restatement of the Gaia hypothesis based on Maximum Entropy approaches to the Earth system. Refuting conceptions of Gaia as a homeostatic system, Gaia is seen as a non-equilibrium thermodynamic system which continuously moves away from equilibrium, driven by maximum entropy production which materializes in hierarchically coupled mechanisms of energetic flows via dissipation and physical work. I propose to relate this view with Kauffman's 'Fourth Law of Thermodynamics', which I interprete as a proposition about the accumulation of information in evolutionary processes. Then, beyond its use in the Kleidon model, the concept of physical work is expanded to including work directed at the capacity to work: I offer a twofold specification of Kauffman's concept of an 'autonomous agent', one as a 'self-referential heat engine', and the other in terms of physiosemeiosis, which is a naturalized application of Peirce's theory of signs emerging from recent biosemiotic research. I argue that the conjunction of these three theoretical sources, Maximum Entropy, Kauffman's Fourth Law, and physiosemeiosis, allows to show that the Kleidon restatement of the Gaia hypothesis is equivalent to the proposition that the biosphere is a system of generating, processing and storing information, thus directly treating information as a physical phenomenon. I substantiate this argument by proposing a more detailed analysis of the notion of hierarchy in the Kleidon model. In this view, there is a fundamental ontological continuity between the biological processes and the human economy, as both are seen as information processing and entropy producing systems. As with other previous transitions in evolution, the human economy leverages the mechanisms by which Gaia moves further away from equilibrium. This implies that information and natural resources or energy are not substitutes, i.e. the knowledge economy continues to build on the same physical principles as the biosphere, with energy and information being two aspects of the same underlying physical process.

# 1. Introduction: 'Information is physical'

In recent work on the Earth System, there is an increasing interest in establishing the Maximum Entropy Framework as an encompassing paradigm to understand the systemic interactions between the different constituent parts of the system (Kleidon and Lorenz 2005). Originally, the Maximum Entropy Production (MEP) hypothesis has been developed in the context of exploring climate dynamics (Paltridge 2009). This extends on Maximum Entropy method in predicting the behavior of complex systems, which is an estimation technique widely used across many domains (e.g. in economics see Golan 2002). The MaxEnt formalism posits that in order to predict future states of complex systems, it is sufficient to specify the constraints on systems processes, and given the general laws of the domain in question, such as the laws of physics, to assume that the processes will converge towards those system states which have the highest probability to occur. Therefore, the MaxEnt formalism builds on a Bayesian approach to statistical mechanics, in particular, as a general analytical stance towards complex systems (Jaynes 2003).

In the MEP approach, this MaxEnt formalism is given a much stronger, ontological meaning in stating that the systems under scrutiny also maximize entropy in the thermodynamic sense. This conclusion is not warranted by the MaxEnt formalism alone, and even misleading, if only based on it (Jaynes 2003: 350f.). So, the ontological interpretation needs to build on additional physical propositions about the nature of the processes, especially in terms of identifying the causal mechanisms. However, there is a clear and neat connection between MEP and MaxEnt in stating that systems evolve into states which have the highest probability to occur, given a certain space of possible states (Whitfield 2005, 2007). This thermodynamic interpretation has to specify the physical mechanisms of energy transformation and conduction, that drive the evolution towards these states. This is precisely the gist of the recent research on the Earth system.

In this paper, I will present a conceptual framework for the MEP approach in the Earth system sciences that focuses on the discussion and interpretation of Kleidon's (2011) most recent and magistral synthesis of the pertinent literature. Kleidon's approach is also of central significance for ecological economics, because it explicitly refers to the Gaia hypothesis. The original Gaia hypothesis states that the Earth system is a tightly integrated complex system in which life evolves in a way to maintain the biogeochemical, climatic and other physical conditions that are the necessary conditions of its viability (e.g. Lovelock and Margulis 1974, Lovelock 1990). The original Gaia hypothesis has raised a lot of controversies (for a critical review, see Smil 2003: 230f.). In the Kleidon view, the biosphere also plays a central role in the Gaia system, but Kleidon extends the original argument in one substantial way, which is to conceive the Earth system as a hierarchical system of different levels and mechanisms of entropy production. This implies that Gaia is not an equilibrium system mainly building on homeostatic mechanisms, but represents a steady state system out of equilibrium (thus offering an alternative approach to views such as in Margulis 1999, who sees evolution as a sequence of moving equilibria). Even more so, the steady state is seen as moving through time, such that Earth system is driven further away from thermodynamic equilibrium continuously.

This substitution of the homeostasis hypothesis by the hypothesis of an evolving steady state is of substantial interest for ecological economics, because the Gaia hypothesis is frequently invoked in stating positions of 'deep ecology', which see the economy as a subsystem of the ecological system that has disturbing effects on Gaia as homeostatic system (e.g. Faber et al. 1996: 18; Šunde 2008: 179; Eriksson and Andersson 2010: 28). The Kleidon approach raises the opposite question: In which way can we conceive of the human economy as following the same forces that drive the evolving steady state of Gaia? Are the 'disequilibrating' forces at the interface between the economy and ecology just the same forces that drive the interaction between the biosphere and the Earth System in toto?

My paper builds on the Kleidon approach and adds an explicit analysis of the underlying principles of the supposed hierarchy of entropy production. I argue that what is still missing is the analysis of the role of the biosphere and biological evolution beyond the mere identification of chemical and physical causal pathways. I propose that the central phenomenon is the accumulation of semantic information in the structural features of the biosphere, driven by the entropic forces: This is what I call the 'physiosemeiotic' view (Deely 2001, Salthe 2007). This extension can build on a recent proposal by Stuart Kauffman (2000) to expand the set of thermodynamic laws by a 'Fourth Law' that is specific to the biosphere, and which highlights the growth of diversity and complexity via the evolution of information-processing physical entities with certain thermodynamic properties. In the context of ecological economics, there is a direct conceptual link to Ayres' (1994) distinction between thermodynamic information and evolutionary information ('survival relevant information'). I propose that the MEP hypothesis corresponds to Kauffman's Fourth Law, with the former focusing on the energetic flows and entropy production, and the latter on the semantic aspect of the evolutionary processes driven by the entropic forces. Kauffman presents a general conceptualization of the causal principles that undergird the Gaia hierarchy in Kleidon's model, and which I specify as 'self-referential heat engines', i.e. heat engines that apply work on themselves. This approach generalizes over generic biological theories which relate evolution with both an increasing potential of harnessing energy throughputs and accumulating information in relation to the environment, especially in the context of origin-of-life research (see e.g. Lahav et al. 2000, Elitzur 2005, Ben-Jacob et al. 2006; for an approach bridging biology and economics, see Corning 2005). In other words, I claim that entropy production and information accumulation are two sides of the same coin (compare Chaisson 2001, 2005). This view would allow to present an account of Gaia which focuses on one defining characteristic of life, i.e. the processing and storage of information in self-organizing biochemical structures, and, at the same time, would allow for establishing a fundamental ontological continuity between life and the human economy, if the latter, following evolutionary economics, is also seen as a knowledge system in the most general sense.

This approach needs further analytical detail in two respects, which are major tasks of this paper. The first is to make the semantic processes more explicit, which is possible when adopting a physiosemeiotic perspective that generalizes over Peirce's theory of signs (following Stone 2007, henceforth I use the spelling 'semeiosis' to distinguish the Peircian view from other strands of semiotics), which is also pointed out by Kauffman, but not further elaborated on (Kauffman 2000: 111; for similar allusions, see Corning 2005: 374f.). The second is to analyze the notion of hierarchy more deeply, which remains intuitive in the Kleidon approach, and which can also build on the Peircian framework. Extending on earlier contributions (Herrmann-Pillath 2010, Herrmann-Pillath and Salthe 2011, Herrmann-Pillath 2011), I argue that the nature of Gaia can only be fully explained if it is seen as a physical system of evolving, hierarchically structured semantic information, which in turn exerts transformative power on the physical environment. The ultimate theoretical core of this physiosemeiotic view is the thermodynamic approach to information which has been developed especially in the sciences of computing, in which the question of the thermodynamic costs of information processing has been pursued with great rigour, yet still controversially (following Landauer 1961, overview in Maroney 2009). Further, the conceptual connection between information processing, hierarchy and transformative power comes close to the biological notion of niche construction (Odling-Smee et al. 2001), which can in turn be related to biosemiotics, following classical contributions by von Uexkuell.

The paper proceeds as follows. In section 2 I summarize the Kleidon model of Gaia. Section 3 introduces the Kauffman conception of autonomous agents and the related Fourth Law, which I further specify in two ways, namely by introducing the notion of self-referential heat engines and by putting the Kauffman argument into the triadic framework of Peircian semeiosis. Section 4 refers the so conceived Kauffman model back to the Kleidon approach and adds a substantial analytical extension with regard to the concept of hierarchy. In concluding, I continue with a view on the implications of the restated Gaia hypothesis for the analysis of human economy.

## 2. Restatement of the Gaia hypothesis in terms of Kleidon's MEP framework

Kleidon (2011; see also Kleidon 2009, 2010) argues that the basic fact about the Earth System is that it constitutes a non-equilibrium system which is mainly driven by external energy throughputs flowing from solar radiation and by the endogenous processes that resulted from the emergence of life on Earth. To this familiar starting point, he adds two central notions. One is the Maximum Entropy hypothesis as such, the other is the notion of a hierarchy of energetic transmission mechanisms, which amounts to an explicit and quantifiable analysis of causal mechanisms.

The MEP hypothesis states that non-equilibrium thermodynamic processes will approach steady states in which the production of entropy relative to the environment is maximal: What matters is the export of entropy, and not the status of the exporting system. This is important to emphasize because one standard argument against the relevance of thermodynamics for analyzing complex living systems is that their evolution runs against the Second Law (which goes back to the famous Schrödinger 1944 contribution). The MEP hypothesis relates the Second Law to the ensemble of systems and their environments and assumes that the maximization of ensemble entropy is the driving force of the evolution of the exporting system (on the general methodological importance of the ensemble perspective in thermodynamic analysis, see Zeh 2005: 68ff.). Thus, MEP refers to the structural evolution of certain physical systems, which differs fundamentally from the uses of thermodynamics for equilibrium systems, in which the working of the Second Law is tantamount to the loss of structure by means of flows that tend to maximize the probability of realized states, given certain constraints, such as in the case of the diffusion of a gas in a container. The upshot of the MEP argument is that increasing structural complexity does not contradict the Second Law, because it speeds up the production of entropy in the ensemble of system and environment. This implies, that the MEP hypothesis is a most simple and straightforward way to explain the increasing distance between realized states of the Earth System from thermodynamic equilibrium, which is the physical expression of the ongoing evolution of life, including human beings. In other words, the claim is made that the evolution of life is a direct physical expression of the Second Law (compare Annila and Salthe 2010, Mäkelä and Annila 2010).

In the Kleidon approach, the central concept to analyze structural complexity is the notion of hierarchy (for a related view, see Annila and Kuismanen 2009). The Earth system is conceived as a hierarchical system in the following sense. There are certain levels and subsystems of the Earth system in which thermodynamic disequilibrium is generated and maintained by energy throughputs that are exogenous to this level and subsystem. These states of disequilibrium are a potential for work generated by the release, dissipation and transformation of free energy. The physical mechanism generating work is contingent to other levels and subsystems on which thermodynamic forces operate, given their structural features. The result of this causal impact is to drive those other subsystems further away from thermodynamic equilibrium, thus creating a further potential for work. So, the entire process appears to be like a fountain with different levels, in which an initial thermodynamic disequilibrium propagates through a manifold of levels which are connected via energy flows that reflect the causal impact of work outputs which are generated by the receiving levels in the flow of energy.

In other words, if viewed dynamically in the time dimension, the thermodynamic machinery of the Earth System is conceived as a hierarchical system of power flows. In this system, two sources of energy loom especially large. One is solar radiation, the other is photosynthesis, which depends on the former. The latter is central to the entire system, because without photosynthesis, the low thermodynamic efficiency of all other physical processes would only maintain relatively small deviations from thermodynamic equilibrium.

It is important to emphasize that this view focuses on average entropy production per unit of time (a distinction that is also important in the generic thermodynamics of information, see Maroney 2009). The First and the Second Law of thermodynamics imply that there is an upper bound of the energy budget of the entire process, hence also for the potential of entropy production. The MEP hypothesis, as applied on the hierarchical Earth System, claims that the emergence of hierarchy firstly, increases the speed by which energy is dissipated (cf. Vallino 2010), and secondly, increases the variety of mechanisms how this can happen (cf. Niven 2010), given the constraints of physical laws, and given the basic requirement of maintaining the structural viability of the hierarchy. In other words, a sudden entropic collapse of the structure would contradict physical laws, given the current boundary conditions, and would diminish the speed of entropy production in the relevant local segment of the universe. So, the Earth System appears to be a physically viable local hot spot of entropy production.

Empirically, this approach requires to establish causal chains in which transformations across different forms of energy take place. So, the differentiated analysis of material fluxes that are concomitant to the fundamental thermodynamic processes is the core of Kleidon's approach. The basic form is the free (Helmholtz) energy that is generated from heat gradients, both caused by solar radiation and by the the heating in inner Earth. The other forms are mechanical energy, geopotential energy, kinetic energy and chemical energy. The release of these forms of energy results into processes that work towards thermodynamic equilibrium, hence maximize entropy. So, the fundamental heat gradient generates the process of diffusion, mechanical and kinetic energy result into mass transport and friction, or chemical energy results into chemical reactions. All of these processes, taken in isolation, would simply result into the dissipation of free energy. However, in a causally connected Earth system the processes impact on each other, especially in the form of chains, networks and, most important, cycles. In order to analyze these processes, the physical notion of 'work' is central. In this view, the change of entropy of the entire system can be expressed as (Kleidon 2011, eq. 14):

$$dS_{tot} = \frac{dU + dW}{T} \equiv dS_{heat} + dS_{diseq} \qquad (1)$$

which results from the combination of the First Law with the standard definition of entropy. In this formulation, $dS_{diseq}$ corresponds to the work done or exerted upon a system. From this follows, when considering changes through time, that the change of free energy is directly proportional to degree how far disequilibrium is diminishing in a process (Kleidon 2011, eq. 16):

$$\frac{dA}{dt} = \frac{dU}{dt} - T \cdot \frac{dS_{tot}}{dt} = -T \cdot \frac{dS_{diseq}}{dt} \qquad (2)$$

Based on this, one can define the net work that is exerted on a thermodynamic variable as resulting from the interaction between two forces, one is the inflow of power, the other is the dissipation rate of free energy. It is important to realize that for this interaction, time scales matter essentially, that is, the processes resulting in equilibrium depend on the rates of power inputs and the dissipation rates, which in turn are determined by certain structural properties of the processes (e.g. soil erosion as compared to a taifun).

In this framework, Kleidon defines the notion of 'entropic force', which results from the interaction between different subsystems of the Earth system. The entropic force strictly obeys the Second Law in the sense that the resulting dynamics follows the gradient of energy dissipation and hence entropy generation over a certain distance *dx* (Kleidon 2011, eq. 23):

$$dW = \vec{F} \cdot d\vec{x} = TdS \tag{3}$$

And for the rate *D* of the dissipation of free energy (Kleidon 2011, eq. 24):

$$D = T \cdot \frac{dS_{diseq}}{dt} = \vec{F} \cdot \vec{v} \tag{4}$$

Now, for positing the MEP hypothesis, it is essential to establish the correspondence between entropy production and power production, i.e. work generated per unit of time. Power extraction can be measured by Carnot efficiency, however, one has to consider the role of heat generation and competing processes in a system. In this case, the MEP hypothesis corresponds to the Maximum Power extraction state. Thus, the Kleidon approach stands in a long tradition of theories which claim that evolution and natural selection are the causal drivers of the emergence of states of living systems which maximize power fluxes in different forms (Lotka 1922a,b, Vermeij 2004), and which see the Maximum Power theorem as a pillar of ecological analysis (Odum 2007).

The MEP framework is completed in considering the causal interaction between subsystems which results from the impact of power fluxes across subsystems with different modes of energy, such as the impact of heat transport on water flows. For every subsystem, one can formulate a basis descriptor of thermodynamic fluxes which result from the impact of power on the subsystem. The power input drives the system away from equilibrium, which generates another source of power fluxes. Since the adaptive forces driving the subsystem back to equilibrium do not operate instantaneously, the subsystem maintains a state of disequilibrium, depending on the continuous power flux, and therefore can in turn become a continuous source of power inputs for other subsystems. As has been emphasized previously, the time structure of the processes matters much (relative speeds and scales).

The entire argument rests upon one essential assumption, namely that the physical (sub)systems have sufficient degrees of freedom to adapt structurally in a way such that maximum power states can be achieved and maintained. For example, river flows can adapt in a given landscape in terms of branching patterns that maximize flows. Evidently, the Kleidon approach is also close to the literature on the so-called 'constructal law' (Bejan and Lorente 2006, 2010), which states that physical systems with sufficient degrees of freedom will always change in the direction of maximizing access to flows. Degrees of freedom always emerge if there are competing processes which interact in dissipating existing entropic gradients.

Empirically, this approach requires the identification of the main sources of energy that generate entropic forces, which can be quantified to a large degree. The other central task is to identify the structural features of the hierarchy of the Earth system. On first sight, this is also straightforward, as there are two major top-down processes, one resulting from solar radiation, the other from heat generation in inner Earth. So, the radiation balance of the Earth atmosphere creates radiation gradients, which drive temperature gradients. These in turn creates motion in the Earth atmosphere, which influences other material flows, in particular, the hydrologic cycles. Geochemical cycles emerge from the hydrological cycles, and ultimately connect with the biosphere. A similar causal chain can be recognized for inner Earth, where steps include the causal link between mantel convection and crust cycling, for example. The two chains meet at the Earth's surface, and especially in the biosphere. Considering the processes on different levels in isolation, and adding up, Kleidon estimates a global work budget that also allows for putting the relative magnitudes into perspective. So, for example, atmospheric motion is estimated to generate a magnitude of 900TW, whereas the geological forces sum up for only about 40TW. The terrestial productivity of the biosphere is estimated about 150TW, and the role of human activity is scaled about almost 50TW, including primary energy consumption. The pivotal role of the biosphere results from the effects of phototrophic life on Earth, because it creates a source of power flows which is independent from the original hierarchy, and directly tapping radiation as a source for generating chemical energy gradients which can in turn generate power fluxes in the system. Kleidon draws on a number of studies for the different levels involved which suggest that these values approximate the maximum power values, which would also support the assumption that the entire system maximizes entropy production. However, this question is the core issue for confirming or falsifying the entire theory, and the evidence is only supportive, so far, but not conclusive.

Further, this simple hierarchical view needs to be enhanced by considering feedback effects, such as the effects of the hydrological cycles on the upper-level gradients via atmospheric changes. This is particularly true in the context of the Gaia hypothesis which states the central importance of the bottom-up processes, i.e. the role of the biosphere in determining all processes in the Earth System. Therefore, the concept of hierarchy seems to refer only to the causal flows that underlie the energy transformations in the system, and which necessarily start with the exogenous sources, i.e. the sun and inner Earth. Further, the view presented so far is simplified as it does not make the role of the temporal structure explicit, which can be causal force of its own, such as the effects of seasonal variations of radiation impact.

Out of this complex analysis of the thermodynamic hierarchy of planet Earth, Kleidon presents his version of the Gaia hypothesis:

*The Earth system has evolved further and further away from a state of thermodynamic equilibrium in time mostly by the increased generation rate of chemical free energy by life.*

The underlying physical principle is the MEP hypothesis. The MEP hypothesis can be supported in different ways. One is tracing energy flows empirically and to show that these correspond or approximate to MEP states. However, this does not yet provide for a fully-fledged theoretical foundation of the MEP hypothesis. In the current debate over the MEP, the central issue is how the MEP relates with the MaxEnt principle (Martyushev and Seleznev 2006; Dewar 2005, 2009). This reflects the fact that in the corresponding mathematical models, the MaxEnt approach is a way to analyze complex systems by means of a particular statistical approach which is based on Statistical Mechanics. In the argument presented so far, the MEP is exclusively based on phenomenological thermodynamics. In the MaxEnt approach it is argued that complex systems in thermodynamic disequilibrium can be analyzed into a set of constraints and the processes that operate under the constraints. If the observer has only limited information about a system, one way to solve the prediction problem is to assume that the processes will move towards establishing the state with the highest probability, given the constraints. So, the MaxEnt approach is not a physical theory, but a methodology which allows for the adjustment of observers' knowledge in a dynamic prediction task (Jaynes 1957, 1965; Dewar 2009). If the observer fails to establish MaxEnt, the constraints have been misspecified.

There is no dispute about the methodological value of MaxEnt, for example, in estimating complex atmospheric processes. But this does not necessarily establish a theoretical rationale that the Earth System processes, especially the processes in the biosphere, also maximize entropy in the physical sense. Indeed, there is a long history of conceptually flawed uses of the entropy concept in biology and economics (for a critical overview, see Corning 2005: 313ff.). Subsequently, I present an argument why this might be the case. I want to show why the Gaia hypothesis and the union of MaxEnt and MEP are necessary complements. Before tackling this task, one principled methodological point needs emphasis. In the debate about the relation between MaxEnt and MEP, the question looms large whether systems actually stay in MEP steady states, which would be an empirical confirmation. But on the other hand, if a system does not actually stay in the MEP steady state, this cannot be interpreted as a final empirical refutation of the theory. On surface, this relates with the physical fact that time scales of causally interconnected subsystems differ vastly, and disturbances occur recurrently on different size scales. But more fundamentally, there are strong reasons why the underlying ontology of the MEP hypothesis needs to be an indeterministic one (Popper 1982). In an indeterministic world, MEP relates with fundamental propensities of physical processes, which includes the persistence of deviations. Even more fundamentally, the increasing variety of possibilities in an evolving state space has to be regarded as an expression of MEP on its own. Then, it is a question of the ontological position adopted, whether empirical violations of MEP are regarded to be refutations of the theory, which implicitly posits a deterministic view of MEP, or whether they are seen either as expressions of the fact that MEP itself cannot be stated as a law in an indeterministic setting (Ulanowicz 1997, who therefore rejects thermodynamic approaches in ecology), or as reflecting the fact that indeterminism and MEP are necessary correlates. In the latter case, Kleidon's argument is sufficiently strong in empirical terms to corroborate the MEP hypothesis.

## 3. The Fourth Law of Thermodynamics in a physiosemeiotic perspective

### 3.1. Central theoretical propositions

The essence of the problem is how to conceive of the processes in the biosphere in thermodynamic terms. Stuart Kauffman has proposed that these processes follow what he calls a Fourth Law of Thermoydynamics which states, in vernacular, that "the biosphere maximizes the average secular construction of the diversity of autonomous agents and the ways how these agents make a living and propagate further" (Kauffman 2000: 4f.).

Kauffman needs a set of other hypotheses and principles to generate this hypothesis. The first is the concept of an 'autonomous agent'. An autonomous agent is defined as a reproducing system that carries out at least one thermodynamic work cycle. So, the notion of agency is directly related with the notion of thermodynamic disequilibrium: An agent manifests agency because it is able to generate work. A further criterion is that the processes are cyclic, in the sense of a simple thermodynamic engine, such that the state of disequilibrium is reproduced concomitant to the throughput of energy.

This definition presupposes two other concepts. The first is that of an 'autocatalytic cycle'. An autocatalytic cycle is a system of chemical reactions which produce catalysts for other chemical reactions in a way that the reaction chain closes to a cycle. This autocatalytic closure implies that the entire cycle assumes certain collective properties, in particular its structural features and the collective lowering of the energetic thresholds for the reactions involved. All autonomous agents are or build on autocatalytic cycles.

The second concept is 'work'. Kauffman argues that the physical concept of work is underspecified in the sense that, mostly, the structure of constraints is only posited implicitely which determines the distinction between energy dissipation, i.e. heat, and work. For example, a mechanical construction that generates work such as moving a weight is actually designed by an experimenter in a way such that a certain amount of energy is released under the constraints. However, the same is true of any sort of work, which operates under naturally emerged physical constraints (Salthe 2007b). From this follows that 'work' is a physical magnitude that cannot be defined independent from context, that is, the physical structure that receives the energy throughput, which is not a universal regularity in nature, but a singular combination of physical properties specific to particular spacetime coordinates.

Now, a special case of work is work directed at constructing the constraints under which work is generated. This is a defining feature of autonomous agents: Autonomous agents work on their environment in a way that some of the causal effects feed back on the constraints under which they operate, including the structural features of their autocatalytic processes. These feedback cycles are driven by the actions of autonomous agents. The most general feature of these actions is an input-output linkage which connects a measurement with work. Measurement inheres the processes which lead autonomous agents to open up sources of energy that feed their autocatalytic processes. Work output includes actions that are directed at those sources. Thus, e.g. a bacterium detects nutrient gradients in the environment and moves accordingly.

Now, thinking about populations of autonomous agents that compete over a set of possible sources of energy, we can say that a given spatiotemporal distribution of agents and their structural properties reflects the history of the interaction between agents and the environment. This history is non-ergodic for simple mathematical reasons, as the space of possible states increases with the number of possible combinations and partitions of the state space. The number of realized states increases much less than proportionally, such that the probability of returning to the same states is arbitrarily close to zero.

This principle is the most fundamental ontological root of the Fourth Law as stated by Kauffman. He introduces the notion of the 'adjacent possible'. When referring to the set of realized chemical entities in the universe, the adjacent possible is the next reaction step that can result from their interactions and its product. This reaction step can lead towards the creation of a new chemical entity which manifests properties that did not exist previously. In this sense, the evolution of the chemical universe manifests growth in the dimensions of its state space. It is essential to notice that this process is driven by the fundamental thermodynamic forces, because chemical reactions are driven by energy dissipation. Further, it is also evident that the possibility of autocatalysis and cycles is a mover of the emergence of new structural features. Stable autocatalytic cycles can in turn interact with other chemical entities, thus possibly connecting to other higher-level entities and so forth.

This conception of the dimensionality of the chemical state space provides the reference for the diversity of autonomous agents as stated in the Fourth Law. Thus, we can think of two different kinds of diversity. One is the diversity of the dimensions of the state space, the other is the diversity of single instantiations under these dimensions. This distinction corresponds to the distinction between species diversity and individual diversity in biology, for example.

These are the major building blocks that I need for establishing the conceptual connection between the Kauffman approach and the MEP framework. Kauffman relates his approach with his previous work (Kauffman 1993) on the emergence of order in chemical networks, applying certain mathematical models of how connectivity of different kinds emerges and results into pattern formation. From these models he generates the specific assumption that living systems and the systems consisting of living systems operate at the 'edge of chaos'. This refers to a state which is close to the transition to chaotic states, thus just maintaining structural stability while at the same time releasing forces of change at a maximum possible rate, just avoiding collapse into disorganization.

The Kauffman approach still lacks a precise account for the physical equivalence between the concepts used in the Fourth Law (such as 'diversity'), and the established thermodynamic concepts in the first three Laws. I think that this requires a direct physical equivalence between entropy and information which goes beyond the purely mathematical homology between Boltzmann and Shannon entropy. One most simple possibility is offered in Zurek's (1989a,b) treatment of the classical problem of Maxwell's demon, which has loomed large in the related discussions for a century now (overview in Maroney 2009; for Kauffman's discussion of this, see 2000: 87ff.). This is to argue that an information processing entity that interacts with an environment to create ordered states with lower entropy (so, like the demon sorting gas molecules between two departments of a container) will always accumulate information which is singular, that is, being algorithmically irreducible. Zurek argued that the thermodynamic entropy of this representation is a part of the ensemble entropy, so that the Second Law still holds. One way to present this idea is to consider the fact that the demon also needs to read its memory, and that this self-referential process will always need to grow in terms of its own complexity when processing algorithmically incompressible memory items. This would match with Kauffman's notion of the expanding state space, because the algorithmically irreducible memory of the 'demon' represents an entirely new possible state which can causally interact with other states, as in the case of chemical reactions leading to new products. So, if we interpret 'diversity' in Kauffman's Fourth Law in terms of informational complexity, Zurek's argument provides a direct connection with thermodynamics. This directs the attention to the fundamental distinction between what Kauffman (2000: 103) calls 'records', i.e. physical storages of information, or, 'memories', and work as a flow. The role of those physical storages is theoretically essential as the relation between information and entropy only can be established via the entropic costs of the erasure of information during the process of accumulating new information (the 'reset' costs in the theory of reversible computing, see Bennett 2003).

However, this line of reasoning still operates on a highly abstract level. What is lacking is an explicit account of the causal mechanisms that would drive the fundamental thermodynamic processes and, at the same time, the evolution of life. This would also allow for connecting the MEP framework to Kauffman's approach. Currently, I think that the Kauffman framework needs to be extended in two directions in order to fit with the MEP analysis of Gaia. One is relatively straightforward, namely the more detailed conceptualization of the thermodynamic work cycle. The other is to add further detail to the notions of diversity and the state space, which, as we are going to argue, requires the extension into semeiotics.

### 3.2. Self-referential heat engines

Regarding the first, it is important to notice that the notion of autocatalysis is a most general one and does not only include the case of chemical autocatalysis (Maynard Smith and Száthmary 1995, Ulanowicz 1997). All systems in which component processes connect cyclically, with the effect of lowering energetic thresholds in process generation, manifest features of autocatalysis. As such, for example, a cell can be seen as an autocatalytic structure, as well as an ecological system. Given this broad reference of the notion, we can generalize Kauffman's conception as an element in the Gaia framework in the sense of defining an autonomous agent as a self-referential thermodynamic engine along the lines proposed by Garrett (2009) (with 'self-referential' my terminological addition).

As in the Kauffman approach, Garrett distinguishes between an entity and its environment, with the line of separation drawn by a permeable boundary such as a membrane (see fig. 1). The system consisting of the entity and its environment is embedded into a surroundings with lower temperature $T < T_s$ such that heat can be exported. The interface between the entity $x$ and its environment constitutes an energy potential $\Delta G$. The available potential energy in the environment is transformed into unavailable forms by means of transfer of matter across the boundary, which can be measured at the rate $a = \alpha \Delta G$, where $\alpha$ is a system-specific coefficient which reflects the intrinsic physical characteristics of the system that influence the availability of energy throughputs (such as access to flows, or conductivity etc.). The energy made available to the entity is transformed by its operations into either work output $w$ or into heat dissipation $a - w$.

**Figure 1: The self-referential heat engine**

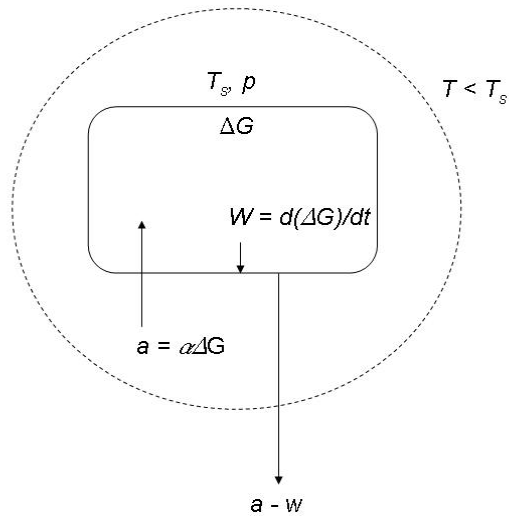

The notion of a self-referential heat engine means that the work output is directed at creating, maintaining and expanding the energy potential $\Delta G$, such that $a = \alpha \Delta G$ and $d\Delta G/dt = w$. The efficiency of the heat engine is defined as $\epsilon = w/a$. Garrett concludes that a system of this kind manifests a growth process, where a rate of return $\eta$ can be defined as (Garrett eq. 2):

$$\frac{da}{dt} = \alpha \frac{d(\Delta G)}{dt} = \alpha \cdot w = \alpha \cdot \varepsilon \cdot a \equiv \eta \cdot a$$

Thus, a self-referential heat engine manifests continuous growth of energy throughputs, depending on the efficiency of the engine.

Based on this argument, we can give a more precise analytical definition of the thermodynamics underlying autonomous agents. At the same time, we can demonstrate how autonomous agents can drive the processes in the Gaia system, in the sense of providing a microfoundation for the analysis of the macro patterns in energy flows. It is important to emphasize that the Garrett model does not imply any particular level of interpretation, that means, the scale and scope of the 'system' in question is not determined. We can analyze a single organism in this way, or a more complex ecosystem with interacting organisms. On all levels, the model describes an endogenous pattern of growth. What is left out of the picture is the fact that all living systems undergo developmental cycles with senescence and decay (Salthe 1993). In this sense, the pure Garrett model may only apply for Gaia in its entirety, with the manifold of developmental processes that form Gaia as contributing forces to the overall growth process.

In this picture, however, growth does not necessarily coincide with MEP. We will argue now that this can be confirmed by analyzing the semeiotic process that is the other defining feature of autonomous agents. The conceptual linkage between Garrett's model and the current biosemiotics is provided by the pivotal role of the boundary, such as a membrane (Hoffmeyer 2000). A boundary establishes a physical correlate to the semantic conception of 'meaning', in the sense that in a self-referential heat engine, all energy throughputs that result into work have a physical format that matches with the internal functioning of the system, and which therefore can pass the system boundary. Physical formats that mismatch with the system trigger processes of entropic decay (wind supports the flight of the bird, the hurricane is a deadly blow). Therefore, all self-referential heat engines can be interpreted as functions. A function is defined as a physical entity which has a causal consequence which is also the cause of the existence of that entity (in more detail, see Herrmann-Pillath 2010, building on Wright 1973). So, a self-referential heat engine produces work that maintains and further expands its difference with the environment, i.e. the energy potential $\Delta G$. Once we conceive of those physical entities as functions, it is straightforward to expand into semeiotic analysis.

### 3.2. A restatement of Kauffman's approach in terms of semeiotics

Kauffman refers to Peirce's theory of signs when discussing the specific workings of autocatalytic systems and autonomous agents (Kauffman 2000: 111ff.). This follows from the observation that any kind of physical measurement involves a categorization. Categorizations imply that in the causal interaction between the system and its environment, only a particular aspect of the environment becomes relevant for the further causal processes. Which aspect, is determined by the internal workings of the system. Thus, the semantic concept of meaning can be directly related with the notion of function (for a related philosophical account in 'teleosemantics', see Millikan 1989, MacDonald and Papineau 2006, Neander 2009). The notion of function also establishes a close connection with cybernetic approaches such as proposed by Corning (2005), which I see as complements, but not alternatives to the view presented here.

Following Herrmann-Pillath and Salthe (2011), I will now proceed in interpreting Kaufman's approach in terms of Peircian semeiotics. I aim at showing that the Fourth Law directly follows from the workings of semeiotic processes, which establishes a physiosemeiotic perspective, or, a naturalization of Peircian semeiotics (Vehkavaara 2002, Herrmann-Pillath 2010). In order to achieve this, we have to introduce the conceptual tool of triadic analysis. Triadic analysis allows for the systematic exploration of two essential process characteristics of autonomous agents, namely categorization and evolution. The term 'triadic' refers to the fundamental Peircian distinction between object, sign and interpretant, and to the confluence of different causal modes in the dynamic unfolding of semeiosis (for a comprehensive overview, see Stone 2007, also Burch 2010).

Semeiosis always involves different modes of causation, efficient, final and formal, in the classical Aristotelian sense. This distinction corresponds to notions of teleonomy that are part of the cybernetic analysis of functions (Corning 2005: 321, Ellis 2008), but makes the semantic dimension of information explicit. It is essential to recognize that the corresponding physical processes interact in the establishment of the fundamental mechanism in the Kauffman approach, namely the autocatalytic cycle, which is the basis of autonomous agents; the autocatalytic cycle is also the fundamental physical mechanisms that relates those different causal modes with energetic throughput (in more detail, see Ulanowicz 1997: 41ff.).

First, every autocatalytic cycle consists of efficient-causal chains of interactions between entities, such as the sequence of singular chemical reactions. Second, once autocatalytic closure is achieved, the property of closure emerges as a property of the collective of entities. In the interaction with the environment, this structure operates as a constraint on the possible variations of the cycles, given a possibility space of many variants which can emerge in a population of entities (such as a chemical solution of different constituent substances). This is downward causation, or formal causation (in a similar vein, see Kauffman 2000: 128ff.). Thirdly, in a population of cycles, those cycles with the highest degree of productivity and energetic efficiency will outgrow other possible cycles, thus creating the directedness of the changes in the composition of a population. This reflects final causality. In this triadic framework, the efficient-causal chains are constitutive in the sense of providing the ontological substrate on which the other forces supervene. Final causality does not establish teleology, but refers to the general property of selection to be both selection *of* and selection *for,* hence teleonomy (Stone 2007: 94ff.). In the generic autocatalytic model, this is selection for throughputs of matter and energy, i.e. maximizing chemical potential, which drives the directedness of physical changes. Formal causality is reflected in the emergence of new conceptual levels in ontological hierarchies, such as molecules with irreducible properties (Vemulapalli 2006).

This triadic approach to causality also inheres the semeiotic triad in the narrow sense. Following the naturalistic approach to semeiosis which was developed by Stone (2007) and Robinson and Southgate (2010), I distinguish between three modes of causal interaction, and introduce a different terminology in treating the 'interpretant' as a 'response' $R$. Figure 2 shows the relations between these different modes which should be read as a conceptual map of the underlying ontology, i.e. the three core constituents (object, sign and interpretant/response), the three causal modes and their relation, and the directions of causalities which are driven by the fundamental force of selection that works on both the levels of signs and interpretants (for more detail see Herrmann-Pillath and Salthe 2011). The central idea of Peircian semeiosis is that in considering a living system, hence an autonomous agent in Kauffman's sense, we need to differentiate between an object in the environment and the response (interpretant) of the system: That means, semeiosis relates to physical effects of an environmental input on the output of the system, and these effects are seen as interpreting the environment. The response of the system is in turn embedded into a larger system, in which the response relates with a function. The function relates in turn with a general purpose $P$, which is the formal characteristic of the relevant selection process that guided the emergence of the function. A function, for example, is the harvesting of nutrients by the bacterium, enabled by its reponse to nutrient gradients (signs) in the environment (i.e. movements as interpretants), serving to maintain its energetic metabolism, which relates with the general purpose that drives the selection of the specific action patterns of bacteria.

The response is impacted by two different causal mechanisms. One is the efficient-causal chain that connects the object impact with the response via the specific physical mechanisms $Q$ (the chain $OQR$ in figure 2). The other is the impact of the sign $X$, hence $OXR$. Both channels co-evolve in a process of selection, in which the conjugation of the three modes of causality materializes. This means, in particular, that the $OXR$ channel puts constraints on the range of possible $OQR$ linkages between a particular input and output, and that the way how the environment is represented in the response patterns changes through time (in the original Peircian analysis, this refers to the distinction between immediate, dynamical and final objects). Thus, the coevolution of the two levels results into the generation of information about $O$ in $R$, physically embodied in both causal chains.

**Figure 2: The basic semeiotic triad**

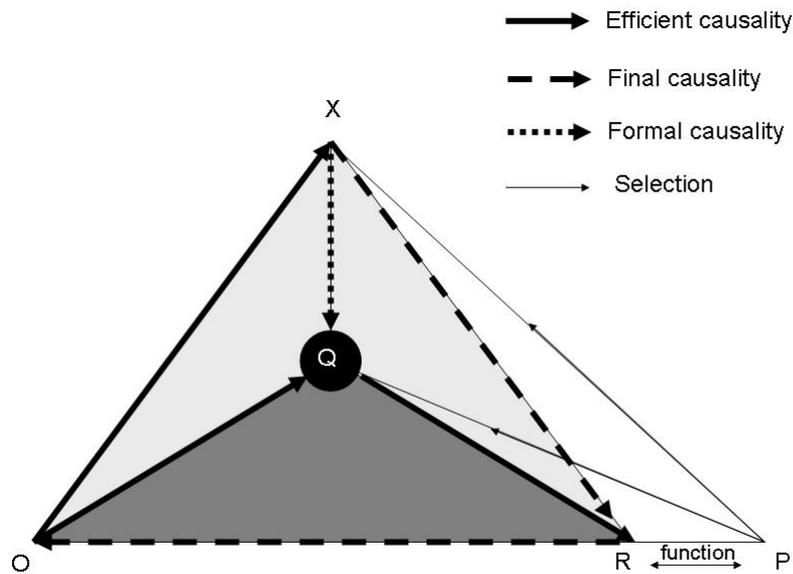

**Figure 3: The semeiotic conception of the autonomous agent**

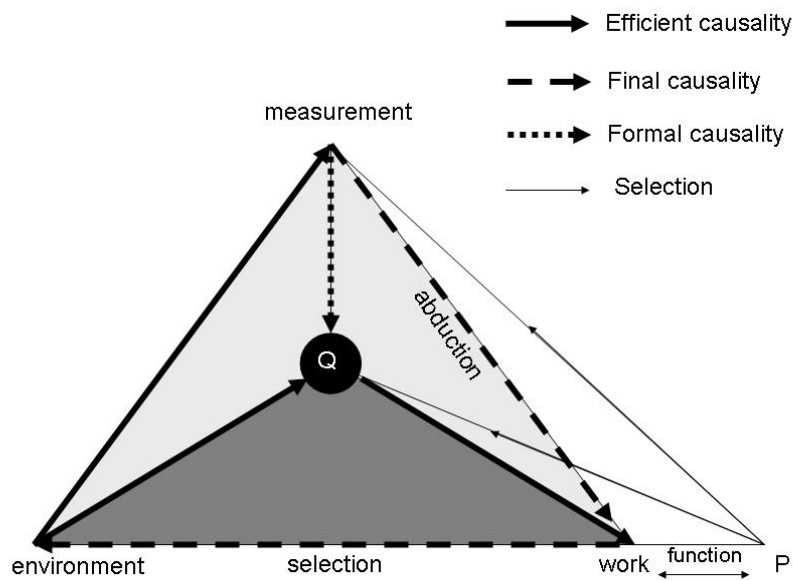

In Kauffman's terminology, the sign corresponds to the measurement, in the sense of assigning an object to a class of measures, which therefore turns into a most simple form of representing the environment to the autonomous agent. Measurements define ranges and scopes of causal interactions between system and environment, and relate with certain classes of responses (compare Dretske 1981). In particular, measurements introduce a gap between the set of all possibly relevant microstates of the environment and the macrostates as represented by signs that become causally relevant for the response *R*. So, we end up with two versions of the original semeiotic triangle, with one directly adapting to the Kauffman terminology (figure 3). In the latter version, we can immediately recognize that the notions of work and of function are complementary in the sense of 'tasks', which has also been proposed by Kauffman (2000: 104f.).

I call this kind of semeiotic structure a 'physical inference device' (for a general definition, see Wolpert 2008). Autonomous agents are physical inference devices, because they are able to infer certain properties of the environment from measurements. Now, the central question is how this inference operates. In this context, the Peircian approach is again helpful. This is because any kind of evolutionary process that involves the fundamental mechanism of selection manifests the logical properties of abduction, the logical type of inference that Peirce had added to the classical dualism of induction and deduction. One of the standard approaches in evolutionary epistemology is to equate evolution with the process of falsification (Popper 1972). However, this leaves the question open how the original hypotheses are created, which corresponds to the problem in evolutionary theory how the original forms of life emerge that are then subject to selection (compare Kauffman 1993). Peirce introduced the notion of abduction, which relates the hypotheses to a previous result of accumulating experience (that is, an observation is seen in the light of its most plausible explanation, given previous experience). In contrast, the concept of falsification relies on the logical form of deduction. Deduction refers to the logical deduction of a consequence from a premise; abduction refers to the logical conclusion from the consequence to the premise. The latter step would be logically wrong, unless it is seen in the context of a process of generating and testing hypotheses. The abduction leads to a hypothesis that can be tested subsequently according to Popperian criteria of falsification. This exactly mirrors semeiosis in an evolutionary context. Abduction results from the interpretation of a certain sign *X*, given the records of previous experience. However, whether the resulting prediction holds true, depends on the outcome of the subsequent selection. Hence, abduction does not establish truth, but hypotheses. In this view, the response *R* of the autonomous agent is equal to a hypothesis that is subsequently tested by selection, i.e. is confirmed or falsified, where the fulfilment of the function, relative to a fitness landscape in which selection proceeds, is the criterion of falsification.

However, this view of selection needs to be further specified in one important sense. As we can see from fig. 3, the semeiotic approach implies that the role selection plays in the evolutionary process needs to be reversed in a certain sense, because the environment becomes dependent on the outcome of selection, and cannot be simply seen as an exogenous force that drives selection. This corresponds to the notion of niche creation and selection in biosemiotics and other recent restatements of evolutionary theory. In the triadic framework, the physical causality of selection is always intermediated by the categorizations that emerge from semeiosis, such that the semeiotically intermediated environment, called 'niche' in biology, and the response patterns co-evolve. In Kauffman's approach (2000: 73ff.), this is explicitly recognized when he argues that populations of autonomous agents evolve such that their environments are changed in a way so that the evolved strategies of problem solutions – their response patterns - match the evolving fitness landscape.

I now propose the following. The process of abduction by means of selection can be formalized according to the MaxEnt principle (on the role of MaxEnt in systematizing observed distributions of results of selection, see Frank 2009a,b; for related approaches see Grönholm and Annila 2007 and Dewar and Porté 2008). This is because the MaxEnt principle means that for any kind of predictions over populations of complex phenomena, the most parsimonious way to accumulate information about this domain is to consider the constraints governing the processes in the domain, and then to assume that the domain will evolve towards a state which is the most probable one, thus maximizing entropy in the sense of statistical mechanics. If the prediction fails, this is an indicator that the constraints have been misspecified. This means, the MaxEnt principle drastically reduces the information requirements of predictions (Dewar 2009). This can be stated in the micro/macro distinction, such that the direct conceptual equivalence with the framework of Statistical Mechanics comes to the fore: Signs and response are two causally linked macrostates, whereas the causal chain *OQR* operates on the level of the microstates. Given the constraints to process information about all microstates, the MaxEnt inference mechanism is the optimal one. It is straightforward to relate this to the triadic framework. Then, we can say that the evolutionary process results into a mapping between between properties of the population of autonomous agents and the environment, such that the former reflect the constraints operating on the environment in a way that simultaneously the diversity of non-essential variations is maximized. In this formulation, it is straightforward to see how the MaxEnt formalism and the evolutionary mechanisms correspond (fig. 4).

**Figure 4: MaxEnt in the triadic framework**

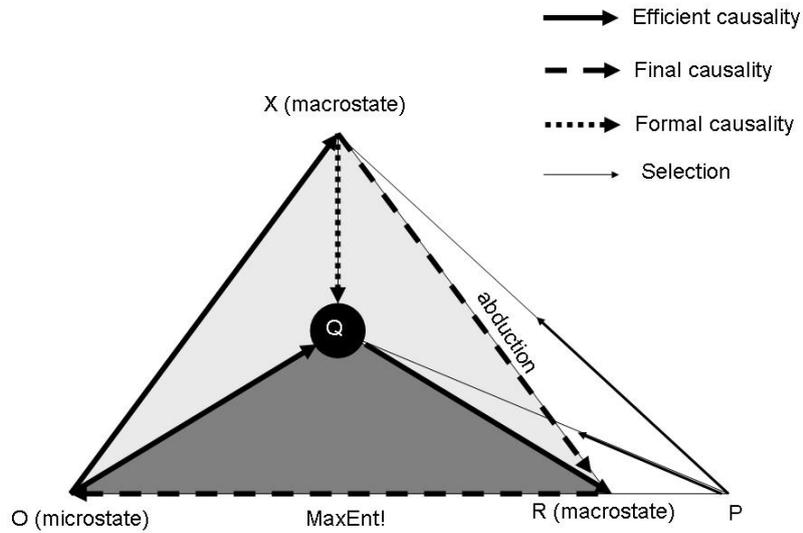

**Figure 5: Physiosemiosis of biological evolution**

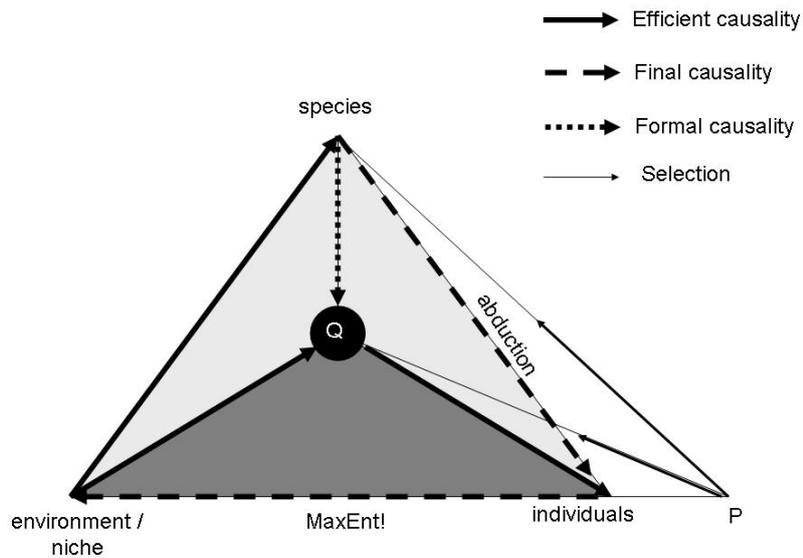

So, the triadic framework enables us to deduce Kauffman's Fourth Law as a consequence of the general properties of the evolutionary process, conceived of as a process of accumulating information: The process of abduction is the force driving the different possibilities of categorizing the state space of the environment, and the MaxEnt process maximizes the diversity of actions, given a certain categorization. So, evolution both increases the dimensionality of the state space, and the number of individual actualizations, which vindicates the Fourth Law. Then we can present a triadic map of the fundamental structure of evolutionary theory, which we call the 'physiosemeiotic' view on evolution (fig. 5). In this view, the species is seen as a set of evolved patterns how to categorize an environment, i.e. is tantamanount to a set of signs. Categorizations elicit individual response patterns, which refers to the population of individuals of a species. These patterns manifest maximum diversity, following MaxEnt, given the constraints elicited by the working of formal causality in the triadic scheme. Ultimately, environment and species co-evolve via the process of niche construction.

### 3.3. Completing the physiosemeiotic synthesis: The 'Fourth Law' restated

If we put the results of this and the previous section together, we can further state that the process of information accumulation is driven by the fundamental energetic dynamics of autonomous agents: physical inference devices are self-referential heat engines. Whereas the abduction cum MaxEnt logic corresponds to the Statistical Mechanics view, the energetic view corresponds to phenomenological thermodynamics. Then we can see that the process of information accumulation corresponds to the increasing potential for work, in the sense of the Maximum Power theorem seminally proposed by Lotka. The possibility of work emerges from the physiosemeiotic process that establishes a complex causal connection between macrostates (signs and responses qua functions) that supervene on the microstates, i.e. efficient causality. Work and information are physical correlates.

**Figure 6: The physiosemeiotic process: The 'Fourth Law'**

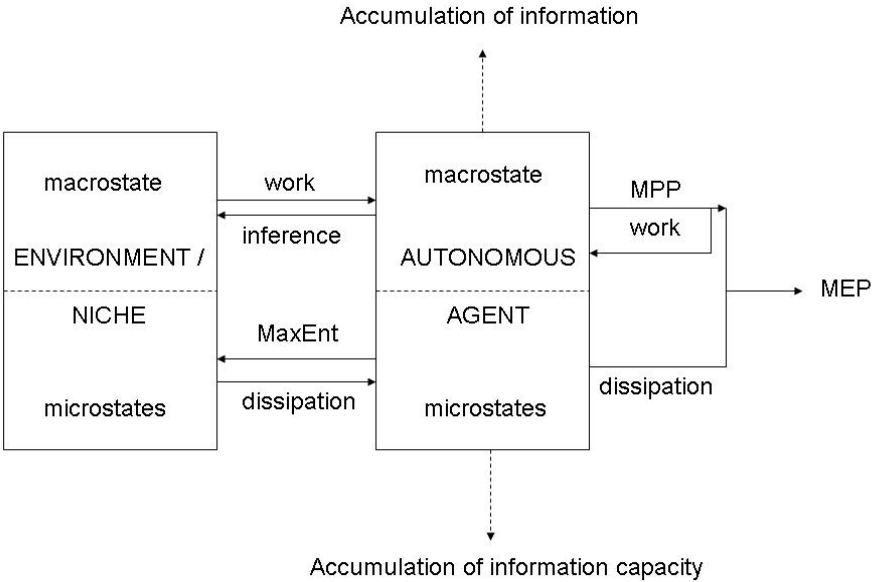

The role of work is twofold. One is that the autonomous agent changes the capacity of the environment to produce work (which is reflected in the Peircian 'interpretation'). That is, a given energy potential is dissipated in a different way when an autonomous agent is present. This is the central piece in developing the general theory of Gaia as proposed by Kleidon. The emergence of autonomous agents changes the physical causality of energy dissipation, i.e. the manifestation of the Second Law. At the same time, this energy throughput, given the information accumulated in the agent (Kauffman's 'records'), enables the agent to exert work on its environment, in order to maintain its functioning. On the population level, this process is evident in two macrotrends. The first is niche construction, such that the potential of the environment is carved out differently by different species, the other is the emergence of species-specific kinds of work, i.e. organismic functionings. At the same time, these functionings concur with a maximum of non-functional variety, which is pure dissipation, at a certain point of time, and given the constraints. But this maximum entropy state also maintains a maximum of information capacity, which is harnessed by the evolutionary process. So, we claim that fig. 6 details the Fourth Law as proposed by Kauffman. Kauffman's Fourth Law, in this formulation, directly relates information-theoretic and thermodynamic approaches to entropy.

The theoretical core of this view, which is also the conceptual bridge to the Kleidon model of Gaia, is the physical concept of work. The accumulation of information means, in physical terms, that new forms of energy transduction and transformation emerge which leverage the existing gradients of dissipation. To give one example: Zehe et al. (2010) present an analysis of soil structures with gradients in water flows, which dissipate kinetic energy. Earthworms can leverage the dissipative gradients, while at the same time exploiting nutrients and other energetic potentials in this environment. The entire system increases the thermodynamic rates in the local environment. Physically, this implies that earthworms represent new qualitative forms of work which are based in their physical structure, which includes the biological storage of information. In my approach, earthworms are physical inference devices which process information about the environment in order to harness energetic throughputs for survival and reproduction, and which thereby maximize entropy production in the soil/earthworm ensemble, their ecological niche.

## 4. Hierarchies in the physiosemeiosis of Gaia

We will now connect the physiosemeiotic argument with the Kleidon picture. Synthesizing Kauffmann and Kleidon in the triadic framework implies the general hypothesis that the evolution of Gaia is just the dual of the accumulation of information in the biosphere. In other words, the two processes of maximizing entropy and energy throughputs and the process of accumulating information are two sides of the same coin. This view can be vindicated if we reconsider the notion of hierarchy.

In the original Kleidon approach, the notion of hierarchy is intuitive, but also inexact. In fact, Kleidon seems to have two different aspects in mind. One is the causal chain of energetic flows. This chain links up solar radiation with subsequent effects, such as the hydrological fluxes. However, a causal chain does not necessarily imply a hierarchy, in the sense of control mechanisms or the distinction of ontological levels. One cannot say, for example, that solar radiation controls the metabolism of plants, it is just an input that is being controlled. So, the notion of hierarchy seems to flow out from an intuitive idea, which in fact relates with the second aspect, that solar radiation is a more general or encompassing causal force than the more specific ones. But this idea is also not exactly related to scale, as it might appear on first sight, because the causal mechanisms working, for example, in the case of radiation can be broken down to the molecular level and below, with the aggregate of these effects not necessarily being more than the sum of its parts. An exact notion of scale would have to refer to the scope and reach of certain functions. Again, this leads to an inverted view of the intuitive hierarchy, because increasing scope and reach of functions are emergent properties of the evolutionary processes. So, Kleidon's use of the term 'hierarchy' seems to refer implicitly to the conceptual order of the physical theories and propositions that describe the Earth system, so that 'radiation' is a more general concept than a local ecosystem, also related with physical processes of larger reach (such as hydrologic cycles). However, considering the underlying causal mechanisms, one cannot speak of a hierarchy in the ontological sense, because all efficient-causal changes are located in particular segments of time and space.

A more precise notion of hierarchy refers to two different uses of the term. One refers to control hierarchies, in which lower level processes are constrained and controlled by higher-level processes. This equates with a hierarchy of functions. The other use refers to semantic hierachies, which are reflected in nested conceptual structures used by observers who describe them, and which reflect the result of an historical process of emerging differentiations in a particular domain. So, I distinguish between two kind of hierarchies, the compositional or scale hierarchy and the subsumptive hierarchy (following Salthe 1986, 2002). The two differ fundamentally in relation to the dimensions of time and space, and we can also speak of the distinction between diachronic and synchronic hierarchy (fig. 7).

**Figure 7: Two kinds of hierarchies**

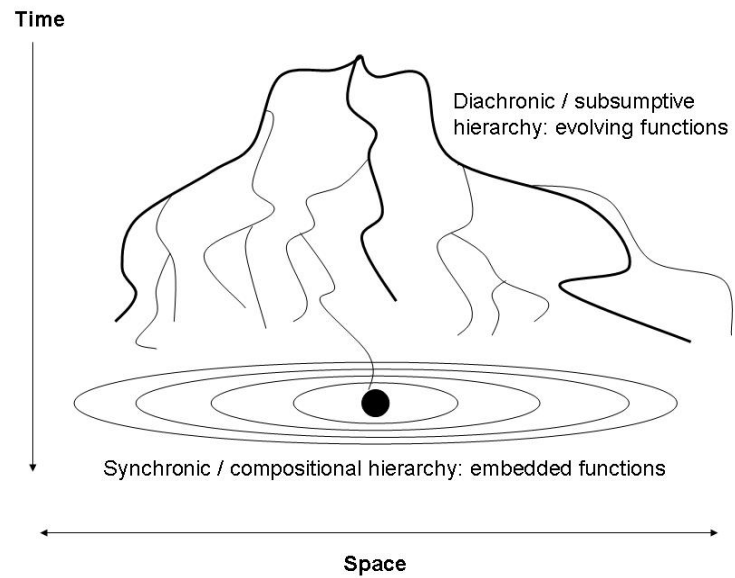

In this view, one problem with the Kleidon approach springs to the eye. The Kleidon picture only focuses on the synchronic hierarchy, and even here it does not further specify the role of functions in the energetic flows. So, only the causal chain flowing from outer to inner circles is part of the picture, but this leaves out the physical mechanisms that ultimately enable the Gaia system to increase the rate of dissipation and to further its distance from thermodynamic equilibrium. Strictly spoken, we can even say that the synchronic hierarchy is irrelevant for analyzing the thermodynamic processes because it leaves the dimension of time apart. Entropic processes occur through time, and hence an explicit treatment of the diachronic hierarchy is necessary. In the triadic framework, Gaia needs to be analyzed as a physiosemeiotic process in which the two dimensions of time and space are considered simultaneously and hence, the two kinds of hierarchies are seen as conjoined phenomena.

This argument can be best illustrated with the primordial process of the emergence of life, as it is conceptualized by Elitzur (Lahav et al. 2001, Elitzur 2005). Elitzur argues that one defining characteristic of life is the emergence of information that increases the potential of its carrier to harness energy in its environment. In this sense, the feature that distinguishes hierarchical levels is the emergence of control information (compare Corning 2005: 360ff.).

Control information integrates a set of efficient-causal mechanisms such that, following Lotka's seminal conjecture, power is maximized, given constraints, under natural selection. The other defining characteristic of life, according to Elitzur, is the emergence of forms, that is, organizational patterns which are autonomous in relation to local conditioned causal forces. Forms can be reflected in conceptual structures, such as the phylogenetic tree of life. Although the evolution of these forms is driven by local forces (variation and selection), the forms coalesce into a non-local structure. This non-local nature of the form is reflected in the evolutionary phenomenon of stasis, that is, the remarkable stability of the characteristics of higher-level biological units such as species (see Gould 2002: 874ff.). This results into another kind of hierarchy, the semantic one, in my previous parlance (for a seminal contribution in this vein, see Riedl 1983).

**Figure 8: Hierarchies in physiosemeiosis**

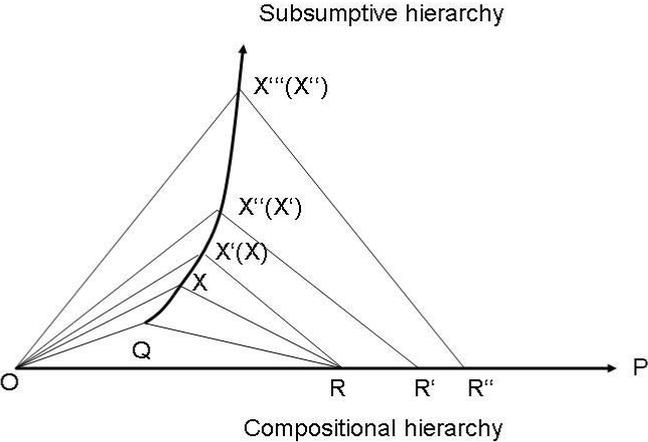

In figure 8 I put the two kinds of hierarchies into the triadic conceptual framework. The point in doing this lies in the fact that only in the triadic framework full account can be given of the role of information in evolution. Following Robinson and Southgate (2010), we can say that the relationship between a particular response *R* and a function is embedded into a compositional relation of responses on different scales, which ultimately connect with a most general purpose, ultimately harnessing energy potentials. For example, if we look at the responses of a bacterium in a Petri dish, we can analyze its response patterns in a very narrow compositional hierarchy, which is limited to the artificial environment created by the experimenter. If we analyze a bacterium in the human intestinal tract, we can analyze its response patterns in functional interdependencies which reach from the immediate response to the internal environment of the human body, and further would include the causal loops which reach from the effects of bacterial actions on the human organism through the functions of the latter onwards to the entire ecological system in which both the bacterium and the human organism are embedded. The energy budget of the bacterium is part and parcel of the compositional hierarchy of ecosystem functions with increasing scale and scope, thus reflecting the original Kleidon view.

However, as this example also shows, the relation between bacterium and human body also reflects a particular state in the evolution of the subsumptive hierarchy. In the context of analyzing the biosphere, the subsumptive hierarchy inheres the genealogical tree that has unfolded since the emergence of life. The subsumptive hierarchy implies an intricate relation between simple and complex semeieotic structures, as reflected in co-evolving patterns of species and their niches. On simple and early stages of evolution, the mapping between signs and functions is direct and close to the ultimate purpose, such as dissipation of energy via chemical reaction cycles. On later stages, the immediate responses of organisms are more and more specified, and therefore in larger distance to the ultimate purpose, but at the same time embedded into the emergent semeiotic hierarchy, in which the phylogenetically older semeiotic structures obtain the role of more general, hence subsumptive concepts that relate more directly with the generic functions closer to the general purpose $P$. So, for example, broader categories in the phylogenetic tree share certain 'Bauplan' features which relate with more generic functions, whereas the defining characteristics of species and subspecies relate with more nuanced features of the niche which had been carved out physiosemeiotically. In fig. 8, this is visualized by the concatenation of signs, which reflects the increasing specificity and complexity of the information processing in evolution, relating semantic categories in the phylogenetic tree with the evolving diversity of niches. For example, in case of the bacterium in the intestinal tract, the symbiotic relation between the bacterium and its host body is regulated by complex systems of signalling which establish and maintain biochemical identities, such that the functional interdependecies can be maintained. This sign system has evolved together with the functional system but cannot be simply reduced to it (on the related issues in analyzing the immune system, see Tauber 2006).

This is not the place to detail this view further, so suffice to state that the peculiar role of the biosphere in the Gaia processes results from two aspects. One is the increasing potential for energy dissipation which emerged in the unfolding of the compositional hierarchy, the other is the energetic flow which is embodied in the unfolding of the subsumptive or diachronic hierarchy through time, and which corresponds to its increasing semantic complexity. In this sense, the distinction of the two hierachies comes close to the economic distinction between stocks and flows, with the flows corresponding to the fulfilment of the functions, and the stocks corresponding to the evolved physical structure of signs. The connection between the two aspects of the same underlying process can be established by referring back to my previous analysis of the Fourth Law. One aspect relates with the production of entropy, which is leveraged by the evolving functional hierarchies, the other aspect is the energy throughput that is embodied in the growing complexity of the information stored in the system, with 'storage' actually referring to the recurrent reproduction and change of that information, which requires the continuous maintainance and expansion of physical structures of embodiment. The second aspect relates with the physical conditions that make the first aspect possible. The two aspects corresponds to the two results of my conceptualization of Kauffman's Fourth Law. The accumulation of information happens in the subsumptive hierarchy, the accumulation of information capacity happens in the compositional hierarchy.

In ecological theory, a framework that comes close to this is the notion of 'ascendency' that has been proposed by Ulanowicz (1997) and Ulanowicz et al. (2009), who also speak programmatically of the 'return of information theory' with reference to ecology. Ascendency is an information-theoretic measure that relates throughputs with the average mutual information in a system, which reflects increasing complexity and determinacy of structure, relative to a space of possible alternative states, hence accumulated information in my sense (e.g. realized network connection in a food web as compared to possible connections, scaled by total throughputs). Ascendency is the scaffold of evolvability, which at the same time depends on what Ulanowicz calls 'overhead', which is unconstrained complexity, hence accumulated information capacity in my sense. This overhead is fed by four main sources. The first is system inputs, hence corresponding to Lotka's conjecture of maximizing energy throughputs. Second is system outputs, which relates with outputs that are functional in the sense of establishing positive feedbacks with the system in question. This corresponds to Garret's conception of self-referential heat engines and Kauffman's notion of work, as adopted in my framework. Third is dissipation, which is energy throughput that is not work in the previous sense, and fourth is redundancy of internal structure, which corresponds to the notion of diversity.

Related to this, the concurrence of the two expressions of the Fourth Law, i.e. the accumulation of information in the subsumptive hierarchy and the accumulation of information capacity in the compositional hierarchy, can be summarily represented in one measure that has been proposed by Chaisson (2001), namely the 'free energy rate intensity' $\Phi_m$ (measured in erg/s/g) which measures the energy throughput per unit of time and mass. This measure is continuously increasing throughout evolution and directly reflects the increasing intensity of information processing in physical entities, or, autonomous agents in Kauffman's sense.

The resulting grand view is summarized in figure 9, which draws on a diagram in Kleidon (2011), but presents some substantial modifications, while simplifying other parts, to avoid cluttering.

**Figure 9: The physiosemeiotic view on Gaia**

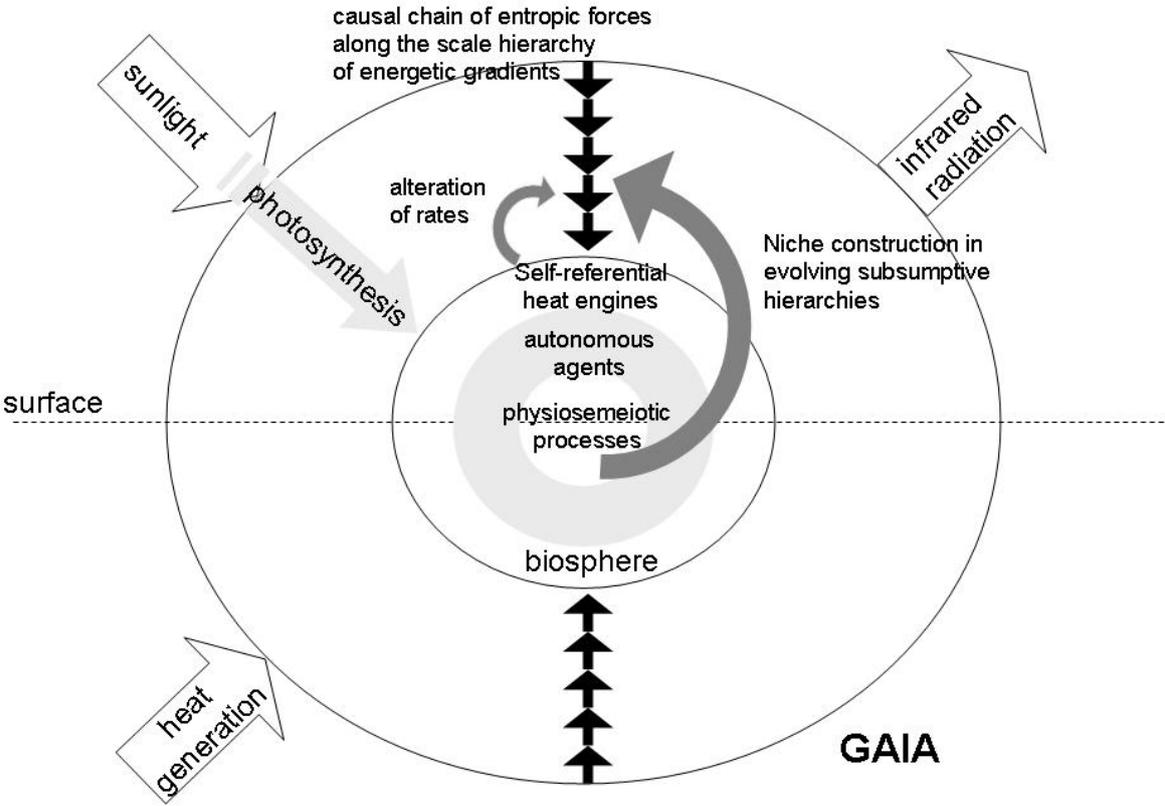

The Gaia system is driven by the energetic flows generated exogenously, i.e. mainly solar radiation and Earth inner heat (subsequently, we leave the latter out of the discussion and the diagram, in order to avoid cluttering). The biosphere is a phenomenon close to the surface of the Earth, but extending beyond in both directions. It stands out as a second source of energetic flows resulting from phototrophic life. The diagram does not present the details about the levels of gradients across which the entropic forces generate work which connects into chains and loops (compare the original, more detailed Kleidon diagram). This is the efficient-causal process that underlies the physiosemeiotic process, thus providing the ultimate justification for extending the notion of semeiosis into physiosemeiosis.

Our diagram differs in opening up the black box of the biosphere. We distinguish between three analytical layers. One is provided by the most general notion of a self-referential heat engine, which catches the Kauffman conception of work as energy dissipated under constraints, and inpacting on those constraints. As we have seen, this notion of a self-referential heat engine is essential to explain the evolutionary directedness of the biospheric changes through time, in the sense of driving Gaia further and further away from thermodynamic equilibrium (Kleidon's definition). Thus, the entire biosphere is a self-referential heat engine. Further, we distinguish between this generic case, which applies for all kinds of ecosystems, and the more specific analytical layer of autonomous agents.

In this context, it is essential to highlight two points. The first is that the biosphere influences the chain of energetic flows (and hence, entropy production) by changing their rates. This is a physical effect that is emergent on the entire biosphere, even though triggered by the single events and action patterns on the constituent individual level of the biosphere (i.e. the organisms). For example, even climate change triggered by human action cannot simply be reduced to the latter alone, but on the entire network of feedback effects in the biosphere. Second, the distinction between autonomous agents and self-referential heat engines refers to the fact that the interaction of autonomous agents triggers emergent phenomena on the ecosystem level which do not imply agency. This distinction is necessary to avoid teleological interpretations of the Gaia hypothesis, but keeps the fundamental teleonomic structure. These emergent properties, in the current context, center around the features of self-referential heat engines, in the sense of intermediating causal chains that redirect work on the constraints under which the biosphere evolves.

Autonomous agents have the properties outlined in the previous sections. Especially, they build on autocatalytic cycles, which are the physical foundation for the emergence of supervening triadic causal mechanism in Gaia. Thus, we argue that a fully-fledged analysis of Gaia needs to move beyond the analysis of efficient-causal chains of entropic forces in order to explain the directedness of changes and the continuous tendency to move away from thermodynamic equilibrium. In the context of autonomous agents, this interplay of causal forces stays at the core of physiosemeiosis, which opens the view on the dual of the entropic flows, namely the emergence and evolution of information that is embodied in the structure of the biosphere. Physiosemiosis is governed by Kauffman's 'Fourth Law' and follows the pattern that was analyzed previously, especially in figure 6.

Now, physiosemiosis also establishes the second major feedback mechanism in the evolution of Gaia, which is the process of niche construction. This results into the twofold hierarchical structure of compositional and subsumptive hierarchies, which changes the nature of the efficient-causal entropic forces fundamentally, as they become embedded into the structure of niches. The concept of niche reflects the principle that has been highlighted by Kauffman, namely that the physical conception of work, as applied in empirical contexts, is fundamentally dependent on the structure of constraints that channels the dissipation of free energy. Therefore, evolving niches continuously transform the physical expression of work.

In sum, analyzing Gaia solely in terms of the causal chains of entropic remains a purely descriptive statement. The more fundamental causal hypothesis of Maximum Entropy Production requires a more powerful explanatory approach, which adds Kauffman's Fourth Law. However, the Fourth Law as such requires explanation, in turn. So, the ultimate explanation is the physiosemiotic one, which shows that the MaxEnt principle and the MEP principle are inherent features of evolution as a process of accumulation information via the emergence of novelty, i.e. the exploration of the adjacent possible in the sense of Kauffman's.

## 5. In lieu of conclusion: Gaia and the human economy

In this grand picture, what is the role for the human economy? The first fundamental insight is that we can no longer speak of a possible contradiction between the human economy and Gaia processes. The human economy extends the information processing mechanisms of evolution, but does not change them in a principled way (compare Jablonka and Lamb 2006). Gaia and the human economy share the feature of driving the Earth System away from thermodynamic equilibrium. Both happen in a unified physiosemeiotic framework, in which the increasing complexity and diversity of information is accompanied by an increasing rate of energetic dissipation and hence entropy production.

This view implies a major rethinking about economic growth, which meets with similar ideas ventilated in the context of the bioeconomics literature (e.g. Vermeij 2009). In a nutshell, economic growth is nothing but another physical expression of the maximum entropy mechanisms that underlie Gaia's evolution away from thermodynamic equilibrium. There is a fundamental ontological continuity between processes in nature and growth of the human techno-ecology. In other words, we can no longer use any notion of 'ecological equilibrium' to identify disturbances caused by human economic activity. From another angle, this implies that the emerging functional systems as human artefacts become part and parcel of Gaia. This fundamental fact is reflected in the data about human energy consumption, which, as has been emphasized by Kleidon, is already more than the energy flows that work in the geological processes. There is one prediction generated from this fundamental idea: The human economy will always increase energy throughputs, in the average. Energy and knowledge are no substitutes, but two aspects of the same underlying physical process.

One pertinent empirical observation relates to the so-called 'rebounce effect', which refers to the observation that technological improvements in energetic efficiency did never result into an absolute reduction of energy throughputs in the economy (Ayres 2005, Ayres et al. 2005, 2008, Warr et al. 2008). This concurrence between economic growth and energy throughputs is a familiar proposition in ecological economics (see e.g. Kümmel 1998, Smil 2008: 335ff.), which could not be supported empirically with the required rigour unless in the recent reformulation presented by Ayres, which refers to exergy throughputs that, depending on technological parameters, transform into throughput of useful work. This restatement of the energy/growth connection exactly matches with Kauffman's Fourth Law framework as outlined in this paper: Growth relates with work, and the external effects of growth relate with dissipation. So, I argue that the Ayres approach is the formal equivalent to the Kleidon model of Gaia in the context of the analysis of economic growth. Clearly, this leads to a fundamental reconsideration of the nature of economic growth; especially, the idea that growth and ecology are opposing principles is obsolete.

In this respect, there is one central question: What is the relation between economic growth and the growth of knowledge? The Gaia framework established in this paper starts out from the fundamental assumption that all information is physical. In the economic setting, this would imply that concepts and measures of economic growth, the growth of knowledge and the growth of energy throughputs are just different ways to approach the same reality. This would overturn common conceptions in many economic theories, where knowledge is just treated as a non-material resource, i.e. a set of 'ideas' (Jones 2002). This is also true for conceptions of the 'noosphere' in philosophical approaches to ecology (for a survey, see Smil 2003: 12f., 266f.), because in the concept of physiosemeiosis, all mental facts are physical facts, too, hence part and parcel of the biosphere. In fact, all knowledge is 'recorded' in physical structures which are being created, maintained, evolve or decay. Can we make sense of this idea in the context of economics?

In fact, energetic throughputs are increasingly harnessed by information processing activities in the economy, as compared to production in the sense of transformation of inputs into outputs. The exact magnitude of these evolving proportions are difficult to assess, because existing statistical categories do not directly match the required analytical categories. Further, there is a strong impact of the evolving international division of labour which makes it difficult to assess trends. For example, the importance of industry as the major consumer of energy is further driven by the rise of emerging economies such as China, which rapidly increase their material living standards. At the same time, this also triggers diverging structural shifts in different countries, such as in the United States, which, partly a mirror image to the Chinese development, specialize even more on information processing industries such as finance. That being said, the following back-on-the-envelope observations may give some reliable hints at the developments.

Following North (1990), we can distinguish between transformation costs and transaction costs in the economy. Transaction costs are the costs of information processing in the broadest sense. The share of transaction costs in GDP is increasing continuously and seems to reach about two thirds in advanced economies today (for a survey of the pertinent literature, see Wang 2003, which goes back to the seminal work by North and Wallis 1986). The methodology is intricate, because, for example, transaction costs in the financial industry actually refer to the information costs of information processing, which in turn relates with transaction costs in other sectors of the economy. However, the essential observation is that the share increases at all: The standard New Institutional Economics approach to transaction costs would predict that economic development implies a decrease of transaction costs in the economy: The transaction cost 'sector' would offer the service of reducing transaction costs. So, the empirical observation contradicts the economic theory. In my framework, this observation is plausible: transaction costs reflect the increasing energetic costs of information processing in the economy. This is different from the costs of organizing market transactions, because the increasing complexity of market transactions is driven by the increasing amount of information that is processed by markets, but what is mainly exogenous to them (such as technological knowledge).

One simple way to identify the energetic costs of information processing would be to assign the measured energy throughput accordingly. The existing data on energy use go via the different kinds of inputs which are measured in physical quantities and then transformed into comparable energy measures (International Energy Agency 2010). We can then assign the sources to the sectors, which are defined according to the standard criteria. The problem is that the sectoral activities also include a large element of information processing (e.g. an industrial robot contains an advanced computer). On the other hand, the services sector certainly has a focus on information processing, but at the same time includes many transformative processes (such as the haircut). In spite of these many inaccuracies, we can state that especially for electricity, the share of industry and transport is decreasing, while total consumption of the 'others' category, which includes agriculture and services, is growing more than proportionally (between 1971 and 2008, total electricity consumption has risen from 439 to 1446 Mtoe, with the 'others' category increasing from 194 to 819,9 Mtoe). The same is true for gas. The picture is different for coal, where industry absolutely dominates, but on a lower absolute level. Oil makes a strong difference, where transport is dominating (growing from 2250 to 3502 Mtoe). However, transport also includes a larger share of information processing costs, not only related to logistics, but also to the kind of transport (e.g. business trips). In the aggregate, the 'others' category hovers around 30 percent between 1971 and 2008. This has to be seen against a secular trend of declining shares of industry in total energy consumption (Smil 2003: 56ff.). The strong role of transport in the recent decades seems to be also related to the evolution of offshoring industries and other forms of distributed industrial processing, which in turn requires advanced information processing. One indicator for this is the extremely high share of value-added that accrues to intermediators such as WalMart and which is shown in the large gap between the producer price e.g. in China and the consumer end price: Transport costs in the narrow sense are only a minor part of these (Head et al. 2010).

Another difficulty is the estimation of the role of human costs of information processing. One alternative approach is just to use the shift in employment across sectors, which involves a rapid increase of the share of services employment in economic development. After all, human beings interact with all other information processing artefacts, and they do much of the information processing in the transformative activities. In this case, it is close to impossible to separate the transformation costs from information processing costs, with the human brain being the most energy-intensive organ in the body. But there are illuminating additional observations. One approach is to look at the energetic costs of buildings, which one can further distinguish across sectors (Smil 2008: 260ff.). The energetic costs of buildings increase rapidly in developed nations and account already for a higher share than industry and transport in the EU, 37 percent as compared to 28 and 32 percent (Pérez-Lombard et al. 2008). Of this, 26 percent is dwellings. According to forecasts, the building sector will be the strongest growing user of energy in the future, which matches with an equal observation on the services sector as the main driver of energy consumption. However, regarding my argument it is difficult to distinguish between information processing related costs and consumption: For example, heating and cooling of dwellings may count differently than of office buildings. Yet, one can argue that one phenomenon clearly reflects the interaction between people and physical structures in the context of information processing, which is urbanization. Treating urbanization as a proxy for the information / energy link would have the advantage to grasp the role of externalities in urban knowledge systems, which is also seen as a major driver of agglomeration. Then, one can argue that the distinction between dwellings and office buildings can be neglected, and the increasing energetic flows through urban structures would be the relevant measure, which can be also calculated as power density (Smil 2008: 308ff.).

Another possible approach is to measure the embodied energy in physical structures that relate with information processing. The difficulty is that all physical structures also embody information, as has been seminally explored by Ayres (1994). This proposal comes close to Chaisson's measure of free energy rate intensity, and after all, the embodied energy of chips by far surpasses those of other industrial goods (Smil 2008: 291). In this context, Odum's (2008) measure of emergy is also pertinent, if broken down to information processing structures and activities.-

From these summary observations I conclude that the human economy reveals the same feature as the Gaia system, that is, the conjoint expansion of information processing and related capacities and energetic throughputs. This is a profound insight, because it questions one assumption that is at least intuitively present in many economic analyses of environmental issues, namely that knowledge and information are substitutes for energy or generally, the use of environmental resources. This assumption is false. The more information accumulates in the human economy, the larger the energetic throughputs become. In the physiosemeiotic view, these two phenomena are actually one.

I conclude. The Gaia hypothesis is one of the much-quoted, yet controversial conceptual frames in ecological economics. Recently, Kleidon has proposed a comprehensive view on the Earth System that allows to reconsider its role. The central idea is to treat Gaia no longer as a homeostatic system, but to analyze it as a non-equilibrium thermodynamic system that approaches steady states which are defined according to the principle of Maximum Entropy Production. I have shown that the central physical concept in analyzing Gaia, work, can be related to Kauffman's proposed 'Fourth Law' in thermodynamics, which can in turn be rendered as a precise hypothesis about the generation and the growth of information in Gaia if we adopt a naturalistic version of Peirce's notion of semeiosis. That is, Gaia is a co-evolving system of signs and correlated functions driven by energetic throughputs and approaching maximum entropy production steady states indeterministically. This view allows for a richer description and analysis of the notion of hierarchy, which is also central in Kleidon's model. In this overarching conception, we can also define a new naturalistic approach to the human economy as an evolving knowledge system.

Ellis, George F. (2008): On the Nature of Causation in Complex Systems, http://www.mth.uct.ac.za/~ellis/Top-down%20Ellis.pdf (accessed January 24, 2010).
Eriksson, Ralf / Anderssson, Jan Otto (2010): Elements of Ecological Economics, Bingdon and New York: Routledge.
Faber, Malte Michael, Manstetten Reiner, Proops John L. R. (1996): Ecological Economics: Concepts and Methods, Edward Elgar.
Frank, Steven A. (2009a): Natural Selection Maximizes Fisher Information, Journal of Evolutionary Biology 22: 231-244.
Frank, Steven A. (2009b): The Common Patterns in Nature, Journal of Evolutionary Biology 22: 1563-1585.
Garrett, Timothy J. (2009): Are There Basic Physical Constraints on Future Anthropogenic Emissions of Carbon Dioxide? Climatic Change, DOI 10.1007/s10584-009-9717-9
Garrett, Timothy J. (2009): Are There Basic Physical Constraints on Future Anthropogenic Emmissions of Carbon Dioxide? Climatic Change, DOI 10.1007/s10584-009-9717-9
Golan, Amos (2002): Information and Entropy Econometrics – Editor's View, Journal of Econometrics 107: 1-15.
Gould, Stephen Jay (2002): The Structure of Evolutionary Theory. Cambridge and London: Belknap.
Grene, Marjorie, ed. (1983): Dimensions of Darwinism. Themes and Counterthemes in Twenti-eth-Century Evolutionary Theory, Cambridge et al. / Paris: Cambridge University Press / Editions de la Maison des Sciences de l'Homme.
Grönholm, Tiia / Annila, Arto (2007): Natural Distribution, Mathematical Biosciences 210: 659-667.
Head, Keith, Jing Ran and Deborah L. Swenson (2010): From Beijing to Bentonville: Do Multinational Retailers Link Markets? NBER Working Paper 16288.
Herrmann-Pillath, Carsten (2010): Entropy, Function, and Evolution: Naturalizing Peircian Semiosis, Entropy 12, no. 2: 197-242, http://www.mdpi.com/1099-4300/12/2/197.
Herrmann-Pillath, Carsten (2011): The Evolutionary Approach to Entropy: Reconciling Georgescu-Roegen's Natural Philosophy with the Maximum Entropy Framework, Ecological Economics 70 (2011) 606–616.
Herrmann-Pillath, Carsten / Salthe, Stanley N. (2011): Triadic Conceptual Structure of the Maximum Entropy Approach to Evolution, forthcoming in: BioSystems, penultimate version: http://arxiv.org/abs/1006.5505
Hoffmeyer, Jesper (2000): The Biology of Signification, Perspectives in Biology and Medicine 43(2): 252-268.
International Energy Agency (2010): Key World Energy Statistics, Paris: IEA.
Jablonka, Eva / Lamb, Marion J. (2006): Evolution in Four Dimensions. Genetic, Epigenetic, Behavioral and Symbolic Variation in the History of Life, Cambridge and London: MIT Press.
Jaynes, E.T. (2003): Probability Theory. The Logic of Science, Cambridge: Cambridge University Press.
Jaynes, Edward T. (1957a): Information Theory and Statistical Mechanics, The Physical Review 106(4): 620-630.
Jaynes, Edward T. (1965): Gibbs vs. Boltzmann Entropies, American Journal of Physics, 33(5): 391-398.
Jones, Charles I. (2002): Sources of U.S. Economic Growth in a World of Ideas, American Economic Review 92(1): 220-239.
Kauffman, Stuart (2000): Investigations, Oxford: Oxford University Press.
Kauffman, Stuart A. (1993): The Origins of Order, Self-Organization and Selection in Evolution, New York/Oxford: Oxford University Press.
Kleidon, Axel (2009): Non-equilibrium Thermodynamics and Maximum Entropy Production in the Earth System: Applications and Implications, Naturwissenschaften 96: 653-677.
Kleidon, Axel (2010): Non-equilibrium Thermodynamics, Maximum Entropy Production and Earth-system evolution, Philosophical Transactions of the Royal Society A, 368: 181-196.